\documentclass[11pt]{article}
\usepackage{jheppub}
\usepackage{amsmath,amssymb,amsfonts,graphicx}
\usepackage{subcaption}

\newcommand{\be}{\begin{equation}}
\newcommand{\ee}{\end{equation}}
\newcommand{\bea}{\begin{eqnarray}}
\newcommand{\eea}{\end{eqnarray}}
\newcommand{\beas}{\begin{eqnarray*}}
\newcommand{\eeas}{\end{eqnarray*}}
\newcommand{\ba}{\begin{array}}
\newcommand{\ea}{\end{array}}

\renewcommand*\d[2][]{%
	\mathrm{d}%
	\ifx\relax#1\relax\else
	\rule{-0.02em}{1.5ex}^{#1}\rule{0.08em}{0ex}\!
	\fi
	#2\,
}

\title{Negative energy enhancement in layered holographic conformal field theories}

\author[]{Alex May,}
\author[]{Petar Simidzija,}
\author[]{Mark Van Raamsdonk}

\affiliation[]{Department of Physics and Astronomy, University of British Columbia,\\
6224 Agricultural Road, Vancouver, B.C.\ V6T 1Z1, Canada.}

\emailAdd{may@phas.ubc.ca}
\emailAdd{psimidzija@phas.ubc.ca}
\emailAdd{mav@phas.ubc.ca}

\abstract{Using a holographic model, we study quantum field theories with a layer of one CFT surrounded by another CFT, on either a periodic or an infinite direction. We study the vacuum energy density in each CFT as a function of the central charges, the thickness of the layer(s), and the properties of the interfaces between the CFTs. The dual spacetimes in the holographic model include two regions separated by a dynamical interface with some tension. For two or more spatial dimensions, we find that a layer of CFT with more degrees of freedom than the surrounding one can have an anomalously large negative vacuum energy density for certain types of interfaces. The negative energy density (or null-energy density in the direction perpendicular to the interface) becomes arbitrarily large for fixed layer width when the tension of the bulk interface approaches a lower critical value. We argue that in cases where we have large negative energy density, we also have an anomalously high transition temperature to the high-temperature thermal state.}


\begin{document}

\maketitle
\newpage
\parskip=10pt

\section{Introduction and motivation}

It is well known that quantum field theories can exhibit various forms of negative energy. 
For example, in a d-dimensional CFT with one periodic spatial direction (labeled by $z$) of width $w$, the energy density takes the form
\be
\label{Tform}
T_{\mu \nu} = {F \over w^d} \eta_{\mu \nu} \qquad T_{zz} = - (d-1) {F \over w^d}
\ee
where $F$ is a dimensionless constant that can be positive for some CFTs. In this case, if $k$ is a null vector whose spatial component is in the $z$ direction, both $T_{00}$ and $T_{k k}$ are negative. The value of $F$ is typically of the order of the central charge of the CFT.\footnote{Throughout this paper, we will use the term ``central charge'' in general dimensions to denote a parameter that measures the number of degrees of freedom. A specific such quantity is the entropy density for the CFT on $\mathbb{R}^{d-1,1}$ in units of the temperature.}

More generally, we can consider such a CFT on $\mathbb{R}^{d-2,1}$ times a strip of width $w$ in the $z$ direction with various choices of boundary physics or couplings between the two sides of the strip. In all cases, assuming that the $(d-2)+1$ dimensional Poincar\'e symmetry is unbroken, the vacuum stress-energy tensor is constrained to take the form (\ref{Tform}). It is interesting to ask whether there is some upper bound on $F$ for a given CFT.\footnote{Alternatively, we can consider the CFT on an open interval of width $w$ and ask if there is a maximum/supremum for $F$ among allowed density matrices with stress-energy tensor of the form \ref{Tform}.}

\begin{figure}
    \centering
    \includegraphics[width=120mm]{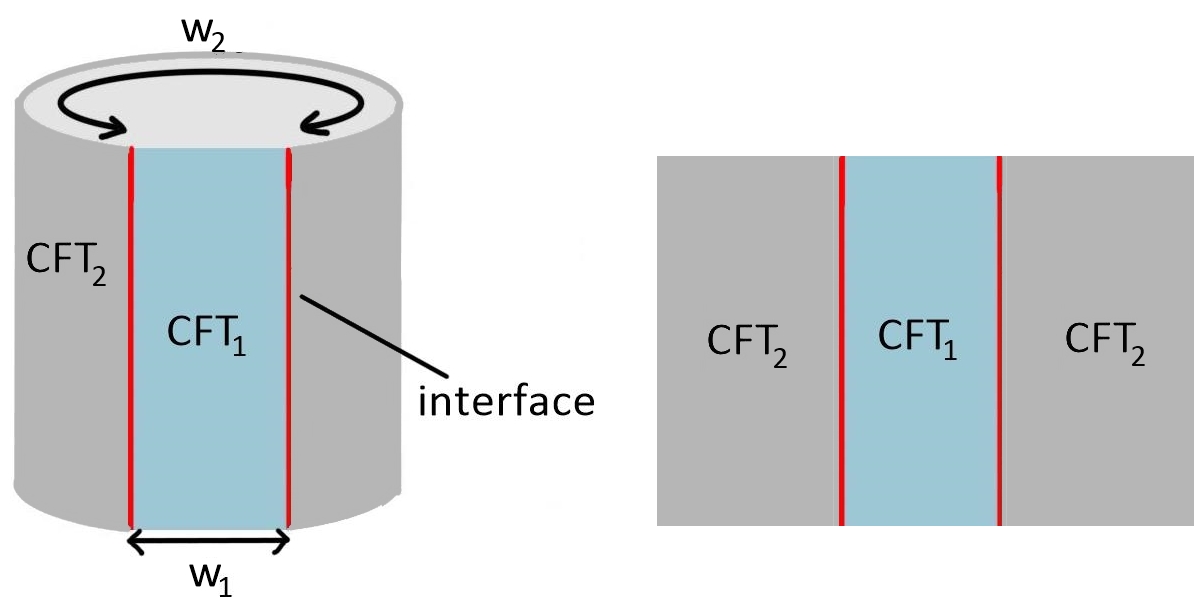}
    \caption{Basic setup: two CFTs, each on $\mathbb{R}^{d-2,1}$ times an interval are glued together on a periodic direction. Right: in the $w_2 \to \infty$ limit, we have a layer of CFT${}_1$ inside CFT${}_2$.}
    \label{fig:setup}
\end{figure}

In this note, we consider the situation where the CFT (which we call CFT${}_1$) is coupled to another CFT (called CFT${}_2$) at the two edges of the strip. The second CFT may also be on a strip, so that the full compact direction is periodic, as shown in Figure \ref{fig:setup}, or it may be extended infinitely in the $\pm z$ directions.

We take both CFTs to be holographic with an Einstein gravity dual, and model the interface between the CFTs via a constant tension domain wall, following \cite{Bachas:2020yxv,Simidzija:2020ukv,Azeyanagi:2007qj,Erdmenger:2014xya}. The parameters in the gravitational description are the AdS lengths in each region and the tension of the domain wall. The gravity solutions dual to the vacuum state of the interface theory are obtained by solving the vacuum Einstein equations together with the Israel junction conditions \cite{Israel1966} at the interface, with the requirement that the asymptotic behavior of the solution matches with the geometry of the CFT setup. From the solutions, we can read off the stress-energy tensor for each CFT via the usual holographic dictionary.

We report the value of the energy density (specifically the parameter $F$) for each CFT as a function of the two AdS lengths, the interface tension, and the relative amount of the compact direction covered by the two CFTs. In CFT language, the AdS lengths and the interface tension in the holographic model correspond to generalized ``central charges'' associated with the two CFTs and the interface. These provide a measure of the number of degrees of freedom for the CFTs and for the interface. The CFT and interface central charges are expected to decrease monotonically under bulk and interface RG flows, respectively \cite{Zamolodchikov:1986gt,Jafferis:2011zi, Klebanov:2011gs, Casini:2012ei,Cardy:1988cwa, Komargodski:2011vj, Nozaki:2012qd, Estes:2014hka, Gaiotto:2014gha, Kobayashi:2018lil,Yamaguchi:2002pa, Takayanagi:2011zk, Fujita:2011fp}. 

\subsubsection*{Enhanced negative energies}

As discussed previously in \cite{Bachas:2020yxv,Simidzija:2020ukv}, the bulk interface tension is constrained to lie within a certain range of positive values in order that the interface can reach the AdS boundary. A striking result of our analysis is that for CFTs of dimension 2+1 or larger, the energy density in the CFT with larger central charge diverges in the limit where the interface tension approaches the lower critical value. Thus, within the context of the holographic model, there is no upper bound for the parameter $F$.

For 1+1 dimensional CFTs, we do not see this effect. Here, $F$ in our model is bounded from above by $c \pi / 6$, the value for the CFT on a periodic direction with antiperiodic boundary conditions for fermions.

The bulk solutions that give rise to large negative energies have an interesting structure, shown in Figure \ref{fig:glued}b. We recall that the solution dual to the vacuum state of a CFT with one periodic direction (assuming anti-periodic boundary conditions for fermions) is the ``AdS-soliton''  \cite{Witten1998a,Horowitz:1998ha} geometry in which the periodic direction smoothly contracts in the bulk. For our solutions with a large negative energy in CFT${}_1$, the region of the bulk associated with CFT${}_1$ is a portion of a multiple cover of the AdS soliton geometry that winds multiple times around the central point but does not include the central point where we would have a conical singularity. It is this winding that allows a negative energy with magnitude larger than a CFT on a periodically identified compact direction with the same width.

The enhancement of negative energies persists when the relative size of the CFT$_2$ interval goes to zero or infinity. In the former case, we end up with CFT${}_1$ on a periodic direction with a particular kind of defect at one location on this periodic direction. In the latter case, we end up with a layer of CFT${}_1$ surrounded by CFT$_2$ of infinite extent as in the right side of Figure \ref{fig:setup}. This latter case is particularly interesting since it could be relevant to real three-dimensional materials.

For specific microscopic CFTs, the allowed interfaces are constrained, and it seems likely that there would be some finite upper bound on $F$ for each specific choice of CFT${}_1$ and CFT${}_2$. It would be interesting to verify this and to understand how large $F$ can be relative to the value corresponding to CFT${}_1$ with a periodic direction.

\subsubsection*{Eternally traversable wormholes and cosmology}

The large negative energies discussed in this work may be exploited in a gravitational context to give interesting solutions. As described in \cite{VanRaamsdonk:2021qgv} (section 4), by turning on gravity in the region with larger central charge, it appears that the negative energy can support a macroscopic eternally traversable planar wormhole. Some tuning of the interface tension is required, but it was argued that this may occur naturally in cases with supersymmetry where the CFT interfaces are each supersymmetric but together break supersymmetry.

\subsubsection*{Related work}

A detailed investigation of such holographic interface CFTs was recently carried out in \cite{Bachas:2021fqo} for both zero and finite temperature in the case of 1+1 dimensional CFTs. Our work complements this paper, in that we mainly focus on the higher-dimensional cases. In particular, the enhancement of negative energy that we find does not appear for 1+1 dimensional CFTs considered in \cite{Bachas:2021fqo}. It would be interesting, following \cite{Bachas:2021fqo} to extend our results to finite temperature in order to work out the phase diagrams more completely in the higher dimensional cases. This is significantly more challenging with two or more spatial directions, since the dual geometries will generally not be portions of the AdS soliton and/or Schwarzschild solutions.

\subsubsection*{Anomalously high transition temperatures?}

While we have not carried out the explicit analysis of the finite temperature case, we argue in the discussion that the properties of the bulk interfaces in the cases with anomalously large vacuum energies would also give rise to anomalously high transition temperatures to the high-temperature phase. It would be interesting to verify this and understand whether it could be relevant to the physics of real-world layered materials.

\subsubsection*{Outline}

The remainder of the paper is structured as follows. In section 2, we describe in more detail the field theory configurations we will consider. In section 3, we review the holographic model for these interface CFTs and the calculation of vacuum energy for one of the holographic CFTs with a single periodic direction. In section 4, we use the holographic model to compute the vacuum energies in the theory with two CFTs arranged as shown in Figure \ref{fig:setup}. In section 5, we suggest a field theory mechanism for the enhancement of negative energies, pointing out that negative Casimir energies in simple free-field calculations are enhanced when infrared modes are removed. We conclude with a brief discussion in section 6. 

\section{Field theory setup}

In this section, we consider quantum field theories built from one or more CFTs, where CFT${}_i$ lives on $R^{d-2,1}$ times an interval of width $w_i$. At the boundaries of these intervals, the CFTs may be joined to another CFT via a conformal interface, or we may have a conformal boundary theory.\footnote{More generally, we could consider the situation where more than two CFTs come together at an interface.} We refer to the direction along the various intervals as $z$ and use Greek indices to refer to the $R^{d-2,1}$ directions.

Assuming that the vacuum state preserves the geometrical $(d-2)+1$ dimensional Poincar\'e symmetry, the stress-energy tensor in each CFT must take the form\footnote{Here, quantum vacuum expectation values are implied.}
\be
T_{\mu \nu} = \eta_{\mu \nu} f(z) \qquad T_{z z} = g(z) \qquad T_{\mu z} = 0 \; .
\ee
The conservation relation $\partial^a T_{ab}=0$ with $b=z$ implies that $g(z)$ is constant. For conformal field theories, the stress-energy tensor must also be traceless, so we have that $(d-1)f = -g$. Finally, by dimensional analysis, we can write
\be
\label{stressgen}
T^{(i)}_{\mu \nu} = \eta_{\mu \nu} {F_i \over w_i^{d}} \qquad T^{(i)}_{z z} = -{(d-1) F_i \over w_i^{d}} \qquad T^{(i)}_{\mu z} = 0 \; .
\ee
The constant $F_i$ can depend only on the dimensionless ratios between $w_i$s, together with our choice of CFTs and interfaces.

Our goal in this paper is to understand the behavior of $F_i$ as a function of the $w_i$s in various cases and in particular to understand how large $F_i$ can be.

\section{Holographic interface CFTs}

We will mostly consider a simple holographic model for the interface CFTs \cite{Bachas:2020yxv,Simidzija:2020ukv}, where the gravity dual of each CFT is Einstein gravity, and the interfaces correspond to domain walls in the bulk between regions of different AdS lengths $L_i$. The gravitational equations are the Einstein equations together with the second Israel junction condition
\be
\label{eq:JC2}
K_{1ab}-K_{2ab} = \kappa h_{ab}
\ee
where $\kappa\equiv 8\pi G_D T/(d-1)$ is defined in terms of the domain wall tension $T$ and $K_{1ab}$ and $K_{2ab}$ represent the extrinsic curvatures for the interface, computed with the bulk metrics on either side, with the normal vector pointing from region 1 to region 2 in each case.

We also assume the first junction condition, which enforces that there is a well-defined metric on the interface equal to the metric induced from the bulk metric on either side,
\be
\label{eq:JC1}
h_{1 ab} - h_{2 ab} = 0 \; .
\ee

\subsection{Asymptotic behavior of the interface}

\begin{figure}
    \centering
    \includegraphics[width=100mm]{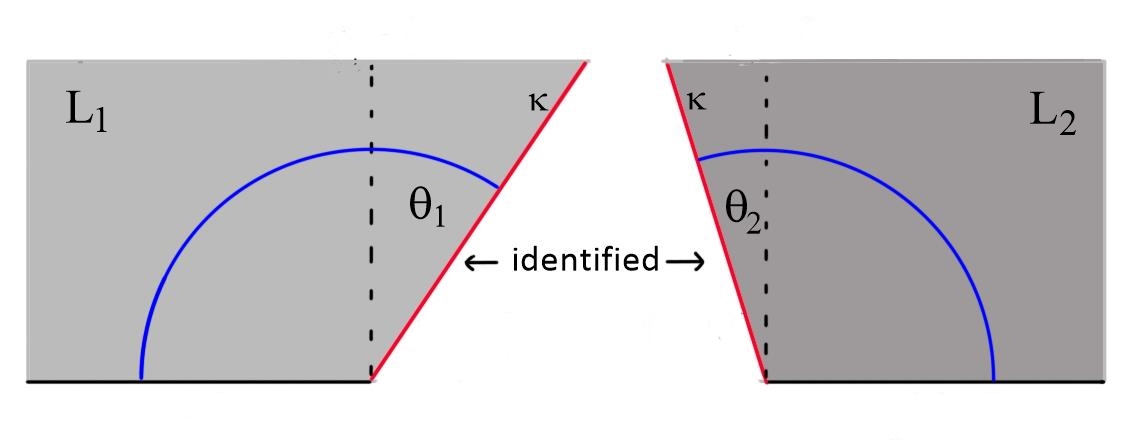}
    \caption{Asymptotic behavior of the domain wall.}
    \label{fig:planar}
\end{figure}

For an interface between CFTs dual to gravitational theories on AdS with AdS lengths $L_1$ and $L_2$, the asymptotic geometry near the interface may be described as a patch of Poincar\'e AdS${}^{d+1}$ with AdS length $L_1$ glued onto a patch of Poincar\'e AdS${}^{d+1}$ with AdS length $L_2$. The domain wall approaches the boundary at a constant Poincar\'e angle in each region, given by \cite{Simidzija:2020ukv}
\begin{align}
     \sin(\theta_1) &= \frac{1}{2} \left(\kappa L_1 + \frac{1}{\kappa L_1} -\frac{L_1}{\kappa L_2^2}\right),\label{eq:rho1}\\
     \sin(\theta_2) &= \frac{1}{2} \left(\kappa L_2 + \frac{1}{\kappa L_2} -\frac{L_2}{\kappa L_1^2}\right)\label{eq:rho2} \; .
\end{align}
Here $\theta_i$ is the angle by which the domain wall deviates from the normal to the AdS boundary in Poincar\'e coordinates. Positive $\theta$ corresponds to an obtuse angle with the AdS boundary, as shown in Figure \ref{fig:planar}.

From these equations, it is found that $\kappa$ must lie within the range
\begin{equation}\label{eq:kappa_cond}
    \kappa \in (\kappa_-, \kappa_+) \equiv \left(\left|{1 \over L_1}-{1 \over L_2}\right|,{1 \over L_1}+{1 \over L_2}\right).
\end{equation}
Domain walls with tensions outside this range cannot reach the asymptotic boundary. 

\subsection{Bulk geometries}

The symmetries of our system imply that the bulk geometry in the region associated with CFT${}_i$ is a portion of the double analytically continued planar Schwarzschild geometry
\be
\label{geom}
ds^2 = L_i^2 f_i(r) dz^2 +  {dr^2 \over f_i(r)} + r^2 d \vec{x}^2
\ee
where
\be
f_i(r) = {r^2 \over L_i^2} - {\mu_i \over r^{d-2}} \; .
\ee
This geometry is also known as the ``AdS soliton'' \cite{Witten1998a,Horowitz:1998ha}. Here, the radial coordinate increases from a minimum value
\be
\label{defrh}
r_{H} = (\mu_i L_i^2)^{1 \over d} \; ,
\ee
the location of the horizon in the analytically continued geometries, to $r=\infty$, the location of the AdS boundary. The coordinates $z$ and $x^\mu$ are defined to be dimensionless; these are related by a constant scaling to the coordinates in the field theory.

When the included region contains the point $r=r_H$, the $z$ coordinate must be periodic with period
\be
\label{defbeta}
\beta = {4 \pi \over d} \mu_i^{-{1 \over d}} L_i^{1 - {2 \over d}} \; 
\ee
to avoid a conical singularity. However, more generally, we can have some region of a multiply wound version of this geometry with a conical excess at $r=r_H$ provided that the region we keep does not include this conical singularity.

\subsection{CFT stress tensor}

For a CFT state described holographically by a portion of the geometry (\ref{geom}), the expectation value of the stress tensor can be read off via the standard holographic dictionary \cite{deHaro2000}. If the field theory coordinates are related to the dimensionless gravity coordinates via
\be
\label{fttograv}
x^\mu_{ft} = \ell x^\mu_{grav}\,, \qquad z_{ft} = \ell z_{grav}\,,
\ee
then the expectation value of the CFT stress tensor is 
\be
\label{stress}
T_{\mu \nu} = {1 \over \ell^d} \eta_{\mu \nu} {\mu \over L^{d-2}} {L^{d-1} \over 16 \pi G} \,\,,\; \qquad T_{zz} = -{d-1 \over \ell^d} {\mu \over L^{d-2}} {L^{d-1} \over 16 \pi G}\,\,.
\ee

\subsection{Single CFT with a periodic direction}

We begin by reviewing the holographic results for the vacuum stress tensor of a single CFT on $R^{d-2,1} \times S^1$ where the $S^1$ has length $w$ \cite{Witten1998a,Horowitz:1998ha}. In this case, one possible dual geometry which is just a periodic identification of Poincar\'e AdS$^{d+1}$. For this geometry, the stress tensor vanishes. 

However, in cases where the $S^1$ is allowed to contract in the bulk geometry (where we have antiperiodic boundary conditions for fermions) we can also have the solution (\ref{geom}). Since the period of the $z$ coordinate in gravity is (\ref{defbeta}) and the field theory period is defined to be $w$, the parameter $\ell$ in (\ref{fttograv}) is 
\be
\ell = w {d \over 4 \pi} \mu^{1 \over d} L^{{2 \over d} -1} \; ,
\ee
so from (\ref{stress}), the stress tensor takes the form (\ref{stressgen}) with
\be
\label{Fbeta}
F = F_\beta \equiv \left({4 \pi \over d} \right)^d {L^{d-1} \over 16 \pi G} \; .
\ee
Up to a numerical factor of order one, this is the central charge of the CFT, providing a measure of the number of degrees of freedom. Thus, in units of the circle size, we have a negative Casimir energy of order the CFT central charge.

It is interesting to ask whether there could be any other solution that gives a lower energy. It was conjectured in \cite{Horowitz:1998ha} that no such solutions exist. In the context of interface theories, we might wonder whether it is possible to lower the energy by having a static bubble of AdS with a different value of the cosmological constant in the interior of the geometry, as shown in Figure \ref{fig:bubble}. In Appendix A, we show that while such solutions exist in some cases (requiring $\kappa$ less than the lower critical value for the domain wall to reach the AdS boundary), they are perturbatively unstable and give solutions with higher energy than the AdS soliton. 

\section{Vacuum energies for interface CFTs}

\begin{figure}
    \centering
    \includegraphics[width=120mm]{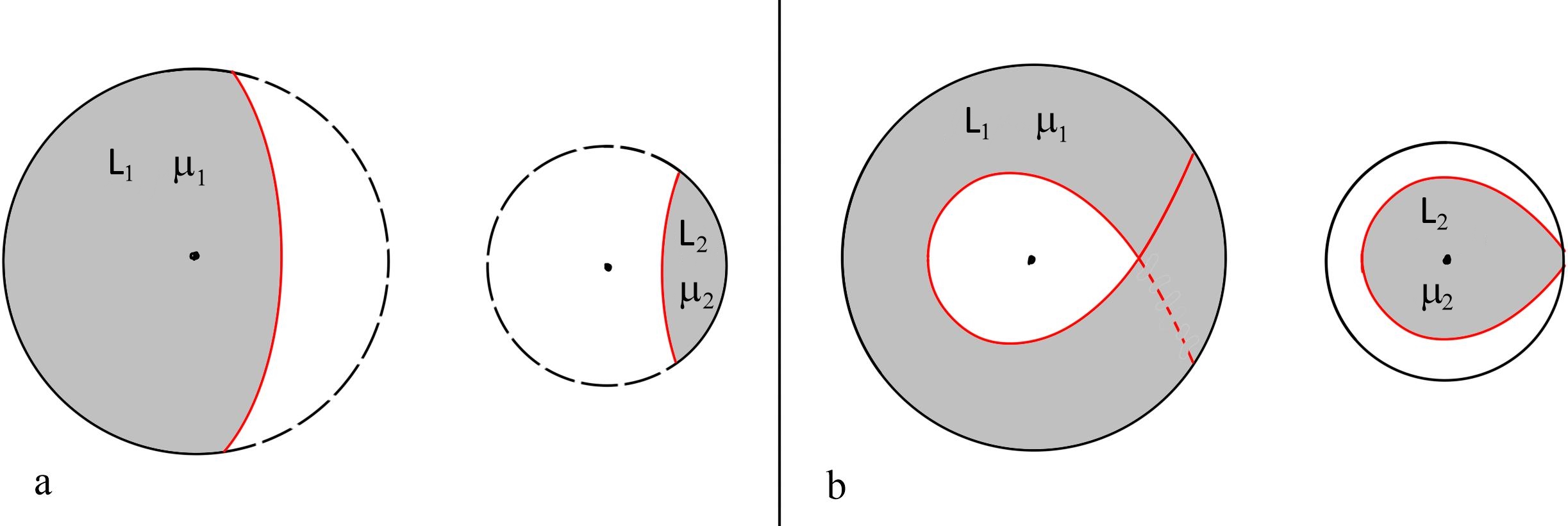}
    \caption{Gravitational solutions for interface CFTs (shaded regions are kept). (a) Regions of AdS soliton solutions dual to the separate CFTs on periodic directions are glued together along an interface. (b) For an interface tension approaching a lower critical value, the outer region is multiply wound relative to the original AdS soliton geometry, leading to an enhanced negative energy for the CFT with more degrees of freedom.}
    \label{fig:glued}
\end{figure}

Now we consider the case where the $S^1$ is divided into two parts, with CFT${}_1$ on an interval of width $w_1$ and CFT${}_2$ on an interval of width $w_2$, as in Figure \ref{fig:setup}. The dual geometries are built from patches of geometries (\ref{geom}) with parameters $(L_1,\mu_1)$ and $(L_2,\mu_2)$ joined along an interface, as shown in Figure \ref{fig:glued}, or Figure \ref{fig:phase2} in the case where $\mu_1 = \mu_2 = 0$.

\subsection{Interface equations from the junction conditions}

Following \cite{Fu2019,Simidzija:2020ukv,Bachas:2021fqo} we can make use of the junction conditions to derive the equations governing the trajectory of the bulk interface.

Without loss of generality, we can identify the coordinates $x^\mu$ in the two patches. The first junction condition then implies that the $r$ coordinates must also match along the interface.

The interface is described by a trajectory $z_1(r)$ in the geometry \ref{geom} dual to CFT${}_1$ and $z_2(r)$ in the region dual to CFT${}_2$; we will consider a region of the interface where the $z$ coordinate in the CFT${}_1$ region increases toward the interface and the $z$ coordinate in the CFT${}_2$ region increases away from the interface.

As explained in more detail in \cite{Simidzija:2020ukv}, the junction conditions
lead to the relation
\be
L_1 f_1 {dz_1 \over ds} + L_2 f_2 {dz_2 \over ds} = \kappa r
\ee
where $s$ is the proper length along the interface. The definition of the proper length parameter $s$ (and the first junction condition that induced geometries on either side of the interface match) gives 
\be
L_i^2 f_i \left({dz_i \over ds}\right)^2 + {1 \over f_i}  \left({dr \over ds}\right)^2 = 1 \; .
\ee
Together, these yield
\be
\label{req}
\left({dr \over ds}\right)^2 - V_{eff}(r) = 0 \qquad  \qquad V_{eff}(r) \equiv f_1 - \left({f_2 - f_1 - \kappa^2 r^2 \over 2 \kappa r} \right)^2 \; ,
\ee
and
\bea
\label{zeqns}
L_1 {dz_1 \over dr} &=& -{1 \over f_1 \sqrt{V_{eff}}} \left({1 \over 2 \kappa r}(f_1 - f_2) + {1 \over 2} \kappa r\right) \cr
L_2 {dz_2 \over dr} &=& {1 \over f_2 \sqrt{V_{eff}}} \left({1 \over 2 \kappa r}(f_2 - f_1) + {1 \over 2} \kappa r\right) \; ,
\eea
where the signs correspond to the conventions described above. The first equation tells us that $r(s)$ will be the trajectory of a particle with zero energy in a potential $-V_{eff}(r)$.

\subsection{Holographic results for the energy density}

We now make use of the interface solutions to provide results for the energy densities in the two-CFT setup of Figure \ref{fig:setup}. 

In this case, it will be convenient to study the ratio 
\be
\label{defE}
E_i \equiv {F_i^{1 \over d} \over F_{\beta}^{1 \over d}}
\ee
of the scale of energy density for each CFT on the strip of width $w$ with the scale of energy density for that CFT on a periodic direction of length $w$. This will be a function of the dimensionless ratio
\be
x \equiv {w_2 \over w_1} \; .
\ee
In our holographic setup, $E_1(x)$ will also depend on the ratio $u = L_2/L_1$ of AdS lengths (related to the ratio of central charges of the two CFTs), and the dimensionless ratio $e = (\kappa - \kappa_-)/(\kappa_+ - \kappa_-) \in [0,1]$, related to the interface central charge. All together, we have a function $E_1(x;u,e)$. The function $E_2(x;u,e)$ for CFT${}_2$ is related to this by
\be
E_2(x;u,e) = E_1(1/x;1/u,e) \; ,
\ee
so we will focus on $E_1$.

\begin{figure}
    \centering
    \includegraphics[width=\linewidth]{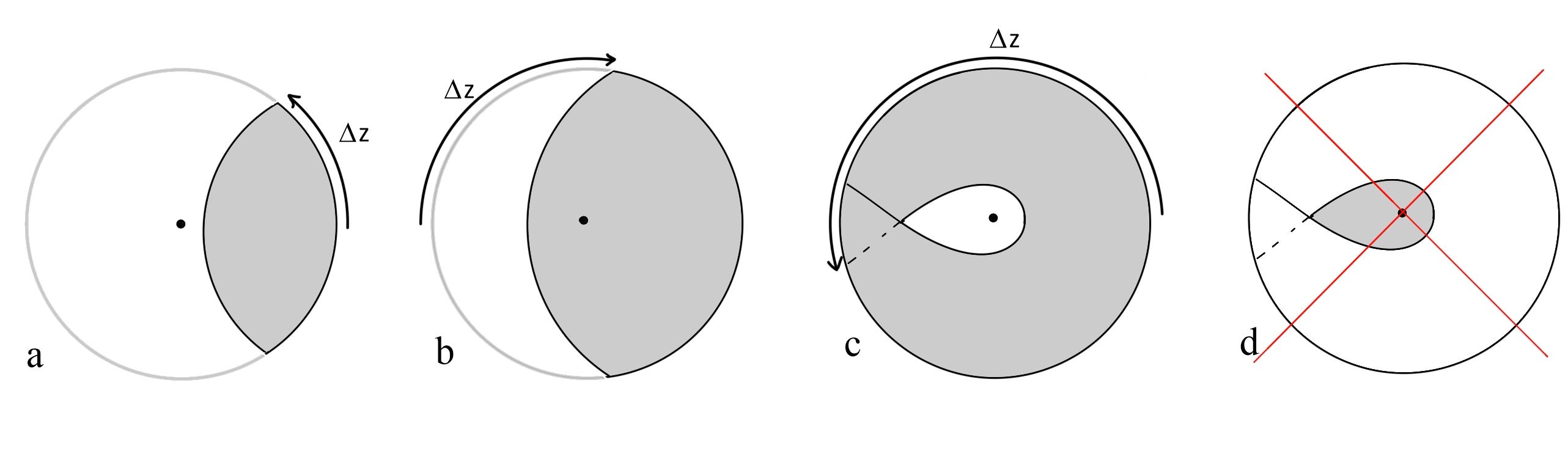}
    \caption{Types of bulk regions associated with CFTs on a strip. (a) The region does not include the Euclidean horizon. (b) The region includes the Euclidean horizon. (c) The region does not include the Euclidean horizon and is multiply wound. (d) Windings less than zero are not allowed.}
    \label{fig:BHoptions}
\end{figure}

In practice, we can study $E_1(x;u,e)$ by first choosing $L_1$, $L_2$ and $\kappa$ corresponding to $u$ and $e$. The parameter $x$ and the energy $E_1$ will then both be determined by the choice of $\mu \equiv \mu_2/\mu_1$ in a way that we now explain.

To find $x$ and $E_1$, it is convenient to define $R_i$ to be the fraction of the asymptotic $z$ coordinate in the solution \ref{geom} covered by the patch associated with CFT${}_i$. This can be larger than one in the case of multiply wound patches as in Figure \ref{fig:BHoptions}c. From the symmetry of the setup, we have that
\be
\label{defR2}
R_2(x;u,e) = R_1 \left({1 \over x}; {1 \over u}, e\right)
\ee
so it suffices to focus on $R_1$.

The parameter $R_1$ is related in a simple way to the change $\Delta z_1$ in $z$ as we move from the point $r = r_0$ where the interface is closest to the center, determined by 
\be
V(r_0) = 0 \; ,
\ee
to the asymptotic boundary. From (\ref{zeqns}), this change is given by
\bea
\label{teq}
\Delta z_1  &=& -{1 \over L_1}\int_{r_0}^\infty {dr \over f_1 \sqrt{V_{eff}}} \left({1 \over 2 \kappa r}(f_1 - f_2) + {1 \over 2} \kappa r\right) .
\eea
The relation between $R_1$ and $\Delta z_1$ depends on whether or not the region that we are keeping includes the center point $r = r_H$, as we can see from Figure \ref{fig:BHoptions}. When the retained region does not include the center, the range of $z_1$ for the included region is  $2 \Delta z_1$ and we have $R_1 = 2 \Delta z_1 / \beta_1$. Otherwise, $\Delta z_1$ is negative, the range of $z_1$ for the included region is $\beta_1 + 2 \Delta z_1$, and we have $R_1 = 1 + 2 \Delta z_1 / \beta_1$.

The retained region includes the center if and only if $dz_1/dr$ is negative as $r$ increases from $r_0$. Since $f_1$ and $V_{eff}$ are always positive, the sign of $dz_1/dr$ is determined by the sign of $f_2 - f_1 - \kappa^2 r^2$ at $r=r_0$.
At the phase boundary in parameter space where this quantity changes sign, the interface crosses the center point, so we have $r_0 = r_H = (\mu_1 L_1^2)^{1/d}$. In this case, the phase boundary simplifies as
\be
f_2(r_0) - f_1(r_0) - \kappa^2 r_0^2 = 0 \qquad \leftrightarrow \qquad  {\mu_2 \over \mu_1} = {L_1^2 \over L_2^2} - \kappa^2 L_1^2 \; .
\ee
We see that $f_2(r_0) - f_1(r_0) - \kappa^2 r^2$ is negative for sufficiently large $\mu_2$, so the condition for our region to include the center is 
\be
{\mu_2 \over \mu_1} > {L_1^2 \over L_2^2} - \kappa^2 L_1^2 \; .
\ee
Thus, we finally have 
\be
\label{defR}
R_1 = \left\{ \ba{ll} {2 \Delta z_1 \over \beta_1} & \qquad {\mu_2 \over \mu_1} < {L_1^2 \over L_2^2} - \kappa^2 L_1^2 \cr
1 + {2 \Delta z_1 \over \beta_1} & \qquad  {\mu_2 \over \mu_1} > {L_1^2 \over L_2^2} - \kappa^2 L_1^2
\ea
\right. 
\ee
In terms of $R_1$ and $R_2$, the ratio $x$ is given by
\be
\label{defx}
x = {R_2 \beta_2 \over R_1 \beta_1} \; ,
\ee
where $\beta$ is defined in (\ref{defbeta}). 

The function $E_1$ defined in (\ref{defE}) that determines the energy of CFT${}_1$ is exactly the same as $R_1$:
\be
E_1 = R_1 \; .
\ee
We can understand this as follows. For an interface solution with some $R_1$ that corresponds to CFT${}_1$ on a strip of width $w_1$, the energy density for this CFT
is the same as that of CFT${}_1$ on a periodic direction of length $w_1/R_1$, since this would have the same local bulk solution and the same identification between bulk coordinates and field theory coordinates. Thus,
\be
T_{ab}[w_1] = T^{periodic}_{ab}[w_1/R_1] = T^{periodic}_{ab}[w_1] R_1^d \;
\ee
where in the last step we have used dimensional analysis. Then
\be
E_1 = \left({T_{ab}[w_1] \over T^{periodic}_{ab}[w_1]} \right)^{1 \over d} = R_1.
\ee

\subsection{Zero energy solutions}
\label{sec:zerosol}

\begin{figure}
    \centering
    \includegraphics[width=80mm]{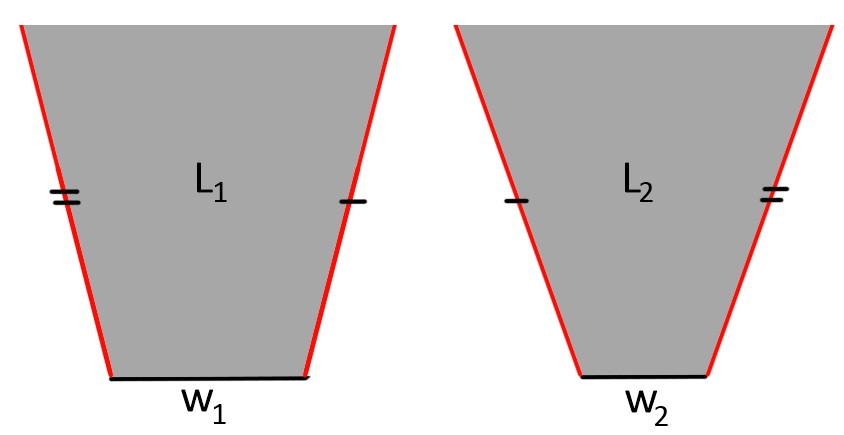}
    \caption{Alternative solutions with zero vacuum energy. The two regions are joined along the domain walls as shown.}
    \label{fig:phase2}
\end{figure}

In addition to the solutions described above, we sometimes have solutions that give zero vacuum energy in each CFT. These are depicted in Figure \ref{fig:phase2}. For these solutions, the local geometry is just Poincar\'e-AdS, with interfaces at fixed angles given by the expressions (\ref{eq:rho2}). 
To avoid self-intersection, these solutions are only allowed when the domain walls in each region tilt away from each other, which requires that $\theta_1$ and $\theta_2$ in (\ref{eq:rho2}) are positive. We find that this will be true when 
\be
\kappa > \sqrt{\left|{1 \over L_1^2} - {1 \over L_2^2} \right|} \; .
\ee
This gives 
\be
e > {1 \over 1 + \sqrt{u+1 \over |u-1|}}.
\ee
Since these solutions have zero energy while the other solutions have negative energy, the zero energy solutions will only correspond to the ground state when solutions of the other type do not exist.

\subsection{Analytic results}

In this section, we provide analytic expressions for $E_1$ and $x$ in terms of the parameters
\be
\mu = \frac{\mu_2}{\mu_1}\,\,,\qquad u = {L_2 \over L_1} \,\,,\qquad e = {\kappa - \kappa_- \over \kappa_+ - \kappa_-} \; .
\ee
From (\ref{defx}) and (\ref{defR2}), the general expression for $x$ is 
\be
\label{resx}
x = \mu^{-{1 \over d}} u^{1 - {2 \over d}} {R_1\left({1 \over u},{1 \over \mu}, e\right) \over R_1\left(u,\mu, e\right)}.
\ee
We now give the result for $E_1 = R_1$ in the cases $u>1$ and $u<1$.

\subsubsection*{Case $u>1$}

For $u > 1$, the relation between $\kappa$ and $e$ is
\be
\kappa = {1 \over L_1} - {1 \over L_2} + {2e \over L_2} \; .
\ee
Defining
\beas
\alpha_0 &=& {1 \over 2}{u^2(1-\mu)^2 \over (u -1 + 2 e)\sqrt{(1-\mu u)^2 + 4 e \mu (u - 1 + e)} + 2(1 + \mu)(1 - e)(u + e) - (1 + u)(1 + u \mu)} \cr
\alpha_1 &=& {1 \over 2} {u^2 (1-\mu) \over (u-1)(u + 2 e) + 2 e^2} \cr
\alpha_2 &=& {(u-1)(u + 2 e) + 2 e^2 \over \sqrt{e(1-e)(u+e)(u-1+e)}}\cr
\eeas
and
\be
{\cal I}_d(a,b,c) = \int_1^\infty y^{-{1 \over d}} dy {y - a \over (y - b) \sqrt{(y-1)(y + c)}}
\ee
we have that
\be
r_0^d = \mu_1 L_1^2 \alpha_0(u,e,\mu)
\ee
and
\be
{2 \Delta z_1 \over \beta_1} = -{1 \over 4 \pi} {\alpha_2 \over \alpha_0^{1 \over d}} {\cal I}_d \left({\alpha_1 \over \alpha_0}, {1 \over \alpha_0}, {\alpha_2^2 \alpha_1^2 \over 4 \alpha_0^2}\right).
\ee
In this case, we have from (\ref{defR})
\be
\label{resR1a}
E_1 = R_1 = -{1 \over 4 \pi} {\alpha_2 \over \alpha_0^{1 \over d}} {\cal I}_d \left({\alpha_1 \over \alpha_0}, {1 \over \alpha_0}, {\alpha_2^2 \alpha_1^2 \over 4 \alpha_0^2}\right) +  \Theta\left[\mu  - \left(1 + {2e \over u} \right) \left({2 \over u} - 1 - {2 e \over u} \right)\right]
\ee
where $\Theta$ is the step function. 

\subsubsection*{Case $u<1$}

For $u<1$, the relationship between $\kappa$ and $e$ is
\be
\kappa = {1 \over L_2} - {1 \over L_1} + {2e \over L_1} \; .
\ee
In this case, ${2 \Delta z_1 \over \beta_1}$ is given by the same expression as above but with
\beas
\alpha_0 &=& {1 \over 2}{u^2(1-\mu)^2 \over (1-u + 2 e u)\sqrt{(1-\mu u)^2 + 4 u e \mu  (1 - u  + u e)} + 2u(1 + \mu)(1 - e)(1 + eu) - (1 + u)(1 + u \mu)} \cr
\alpha_1 &=& {1 \over 2} {u (\mu-1) \over (1-u)(1 - 2 e) - 2 e^2 u} \cr
\alpha_2 &=& -{(1-u)(1 - 2 e) - 2 e^2 u \over \sqrt{e(1-e)(1+eu)(1-u+eu)}}\cr
\eeas
In this case, we have from (\ref{defR})
\be
\label{resR1b}
E_1 = R_1 = -{1 \over 4 \pi} {\alpha_2 \over \alpha_0^{1 \over d}} {\cal I}_d \left({\alpha_1 \over \alpha_0}, {1 \over \alpha_0}, {\alpha_2^2 \alpha_1^2 \over 4 \alpha_0^2}\right) +  \Theta\left[\mu  - \left(1 - 2e\right) \left({2 \over u} - 1 + 2 e\right)\right]
\ee
where $\Theta$ is the step function.

\subsection{Results}

We now describe the results for $E_1(x ; u,e)$ in various cases. 

\begin{figure}
    \centering
    \includegraphics[width=0.5\textwidth]{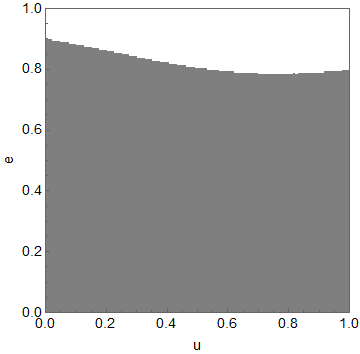}
    \caption{Phase diagram in $d=4$: shaded regions indicate values of e = $(\kappa - \kappa_-)/(\kappa_+ - \kappa_1)$ and $u=L_2/L_1$ for which a connected solution exists and the vacuum energy is negative, while unshaded regions indicate values for which a connected solution does not exist and the vacuum energy is zero.}
    \label{fig:u-e_phase}
\end{figure}

For a given $u$ we find that solutions with a connected interface exist up to some critical value $e^*(u) = e^*(1/u)$. Above this value, there is no choice of $\mu$ for which both $R_1$ and $R_2$ are greater than 0 (i.e. the interface solution always self-intersects in one of the regions). In this case, the dual solution is of the type described in section (\ref{sec:zerosol}), and the energy density in each CFT is zero. The phase diagram showing the values of $e$ for which we have a connected solution with negative energy densities is shown in Figure \ref{fig:u-e_phase}.

\begin{figure}
    \centering
    \begin{subfigure}{0.32\textwidth}
        \centering
        \includegraphics[width=\textwidth]{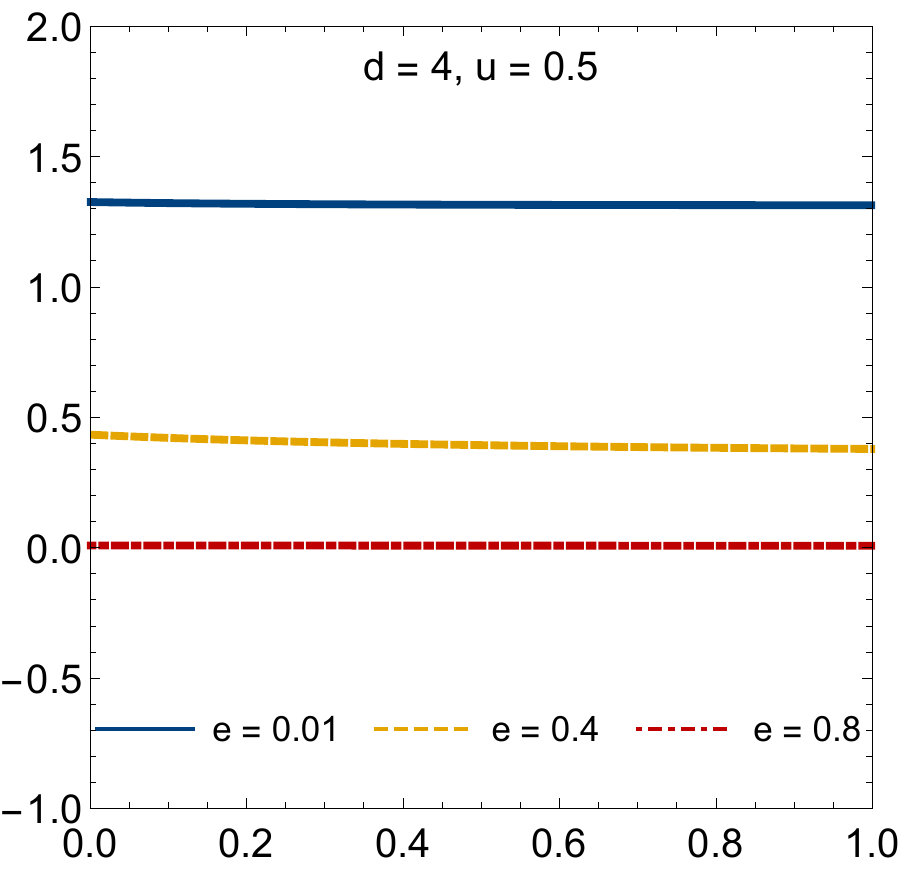}
    \end{subfigure}
    \begin{subfigure}{0.32\textwidth}
        \centering
        \includegraphics[width=\textwidth]{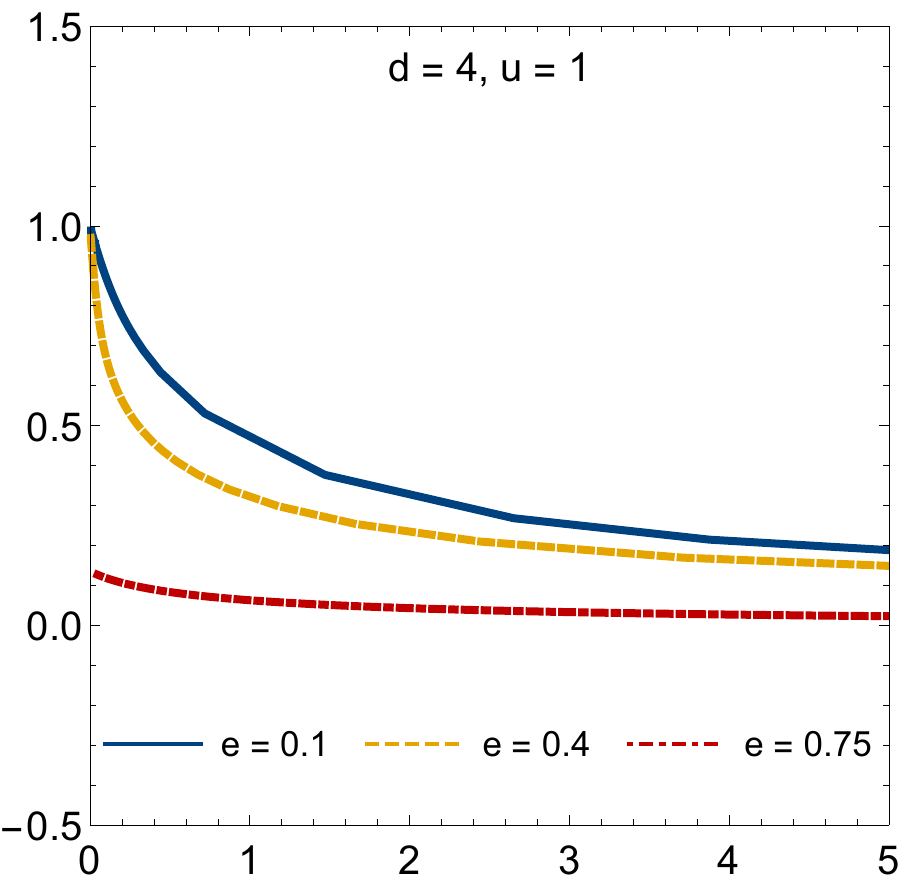}
    \end{subfigure}
    \begin{subfigure}{0.32\textwidth}
        \centering
        \includegraphics[width=\textwidth]{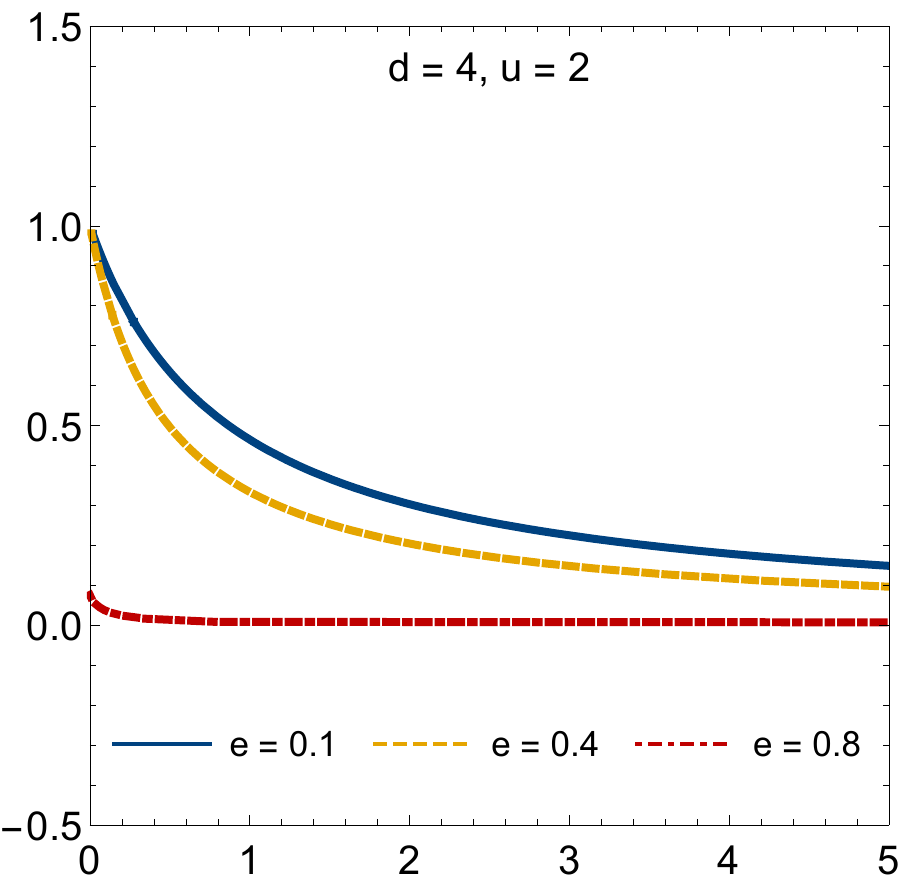}
    \end{subfigure}
    \begin{subfigure}{0.32\textwidth}
        \centering
        \includegraphics[width=\textwidth]{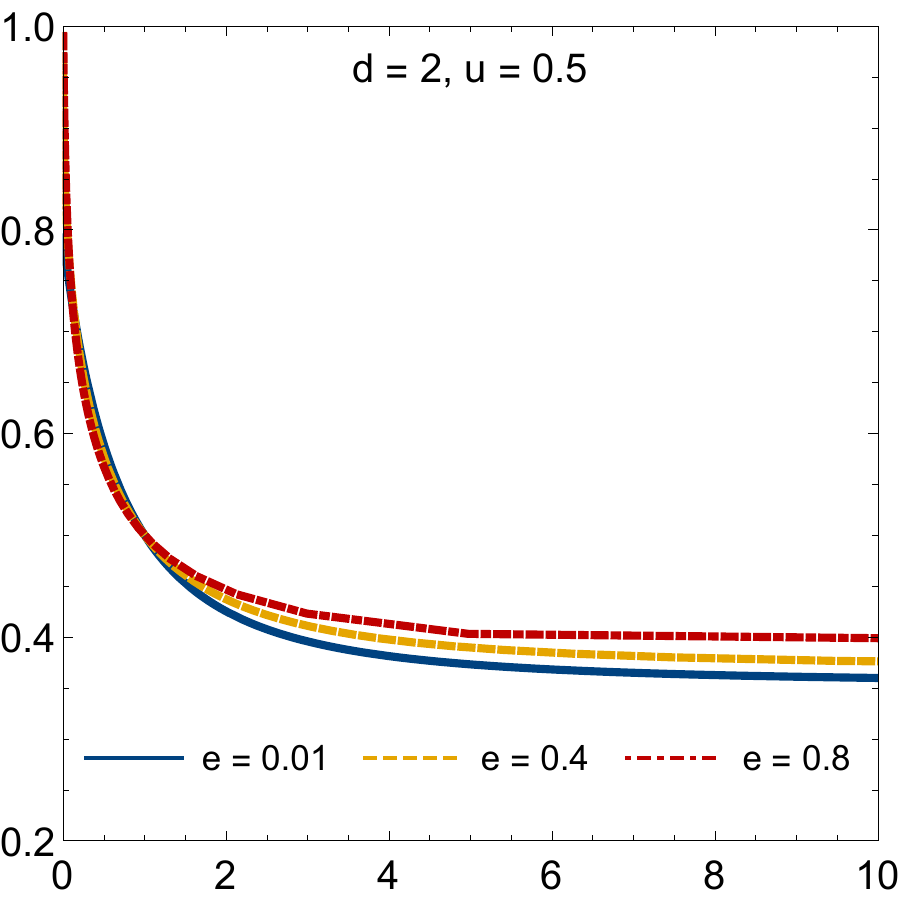}
    \end{subfigure}
    \begin{subfigure}{0.32\textwidth}
        \centering
        \includegraphics[width=\textwidth]{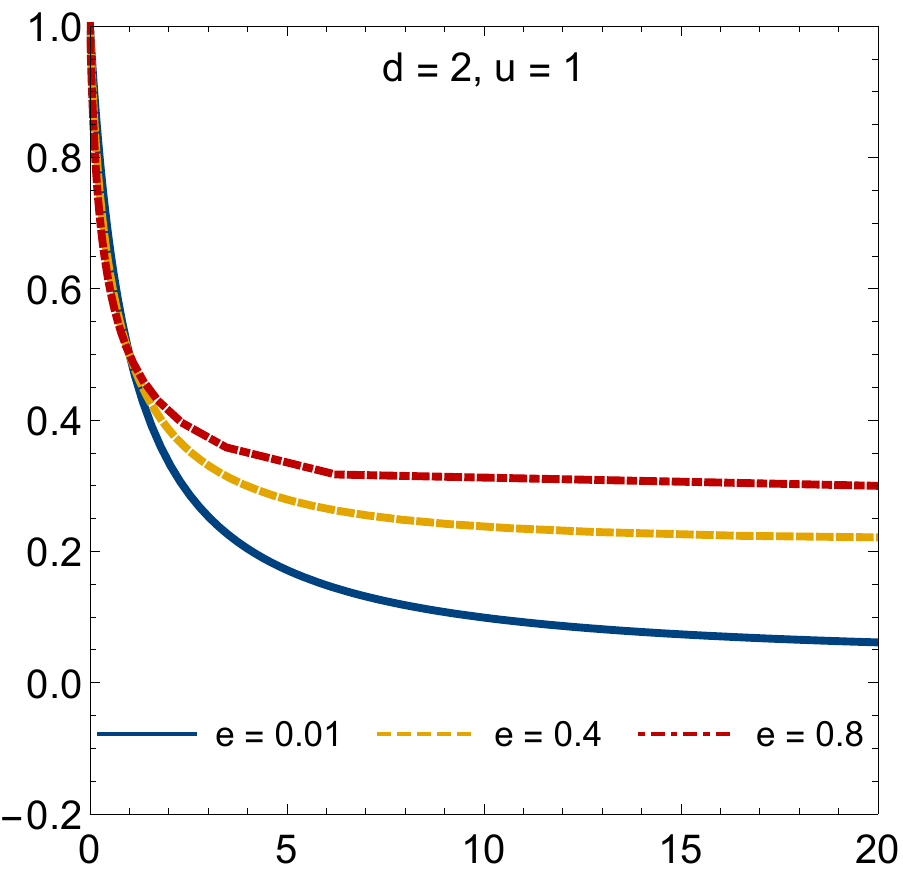}
    \end{subfigure}
    \begin{subfigure}{0.32\textwidth}
        \centering
        \includegraphics[width=\textwidth]{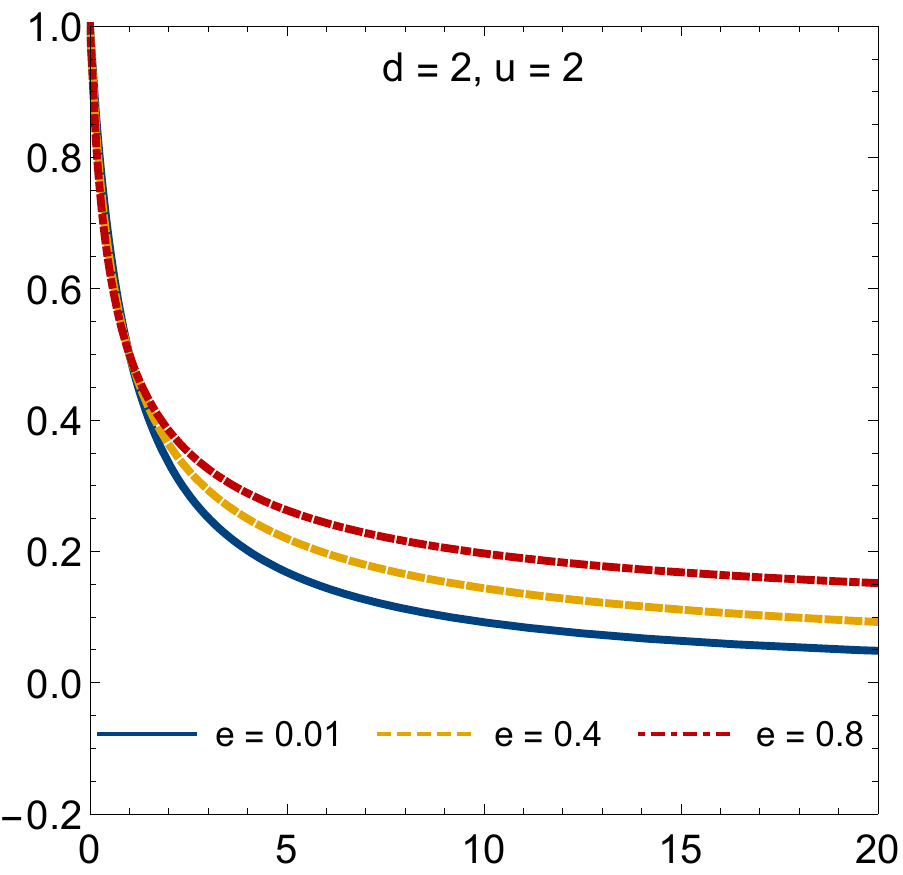}
    \end{subfigure}
    \caption{Plots of $E_1$ versus $x$.}
\label{fig:Evsx}
\end{figure}

Plots of $E_1(x,u,e)$ vs $x$ are given for various $u$ and $e$ below the critical value in Figure \ref{fig:Evsx} for $d=4$ and $d=2$. The plots for $d=3$ are qualitatively similar to those for $d=4$. 

For $u>1$, we find that the value of $E_1$ is always less than 1, approaching $E_1=1$ when $x \to 0$. Physically, this indicates that when CFT${}_2$ has greater central charge than CFT${}_1$, and as the width of the CFT${}_2$ strip goes to zero, the energy density reduces to that of CFT${}_1$ on a periodic direction. In other words, when the interfaces come together, we get a trivial interface. On the other hand, when $u < 1$ (i.e. when CFT${}_2$ has smaller central charge than CFT${}_1$), we do not get a trivial interface when $x \to 0$. This was noted for $d=2$ in \cite{Bachas:2021fqo}. 

\begin{figure}
    \centering
    \includegraphics[width=0.5\textwidth]{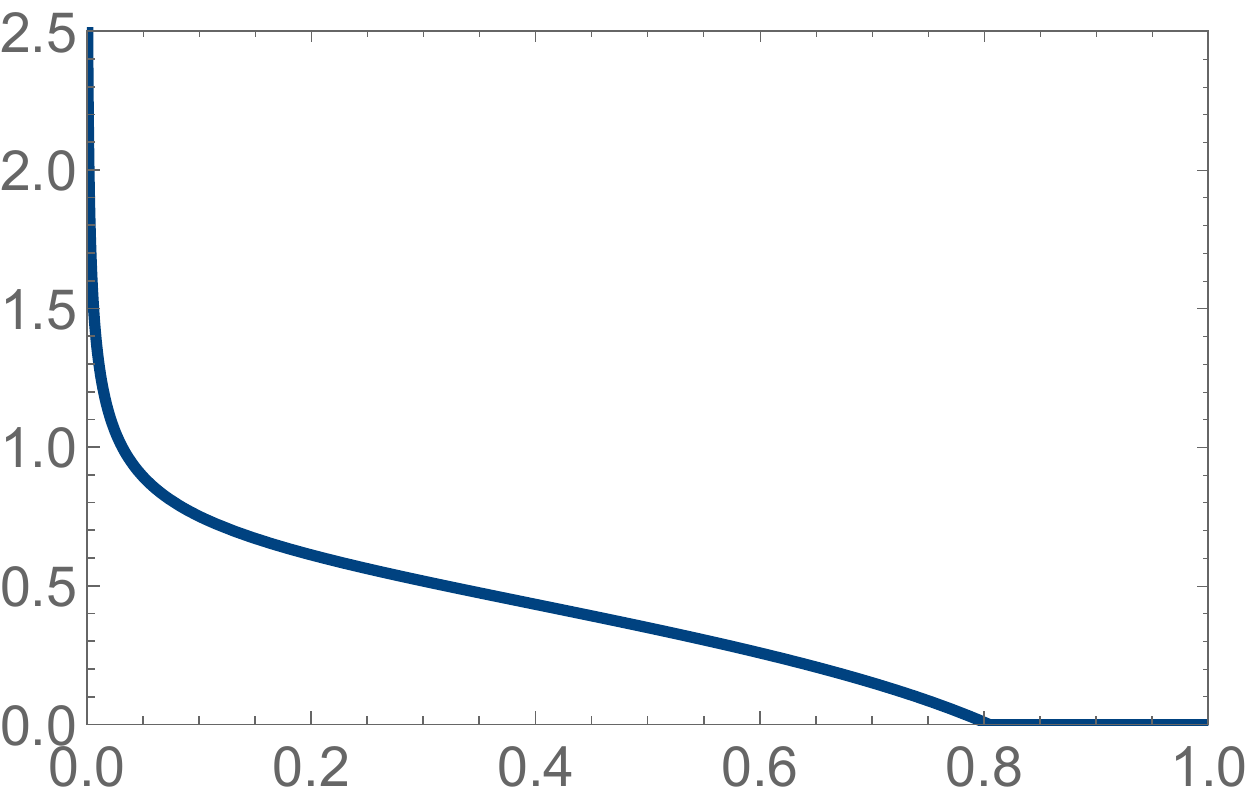}
    \caption{Plot of the maximum value of $E_1$, $E_1^{max}$, as a function of $e$ for $d=4$ and $u=0.5$. Note that $E_1^{max}\rightarrow\infty$ as $e\rightarrow 0$, and $E_1^{max} = 0$ for $e$ larger than some critical value.}
    \label{fig:Emax_vs_e}
\end{figure}

We have a particularly dramatic effect for $d>2$ in the case where $e$ becomes small (i.e. the interface tension approaches the lower critical value). Here we get values $E_1 > 1$, with $E_1$ diverging in the limit $e \to 0$. This means that the magnitude of the negative energy density for CFT${}_1$ in this setting exceeds that of CFT${}_1$ on a periodic direction of the same size, by an amount that becomes arbitrarily large for $e \to 0$. The maximum value of $E_1$ (obtained for $x \to 0$) is plotted as a function of $e$ for $d=4$ and $u=1/2$ in Figure \ref{fig:Emax_vs_e}. 

In the next section, we analyze this interesting case of $u<1$ and small $e$ analytically.

\subsection{Analytic results in the limit $\kappa \to \kappa_-$ for $L_2 < L_1$}

Our numerical results suggest that for $d > 2$, $E_1$ diverges in the limit $\kappa \to \kappa_-$ ($e \to 1$) in the case where we have $L_2 < L_1$ ($u < 1$). Let us now consider this limit analytically. 

For cases where $\mu < 1/u$ (we will see that this is the relevant regime below), we find that in the limit $e \to 0$, 
\beas
\alpha_0 &\to& {1- \mu u \over 4 e} \cr
\alpha_1 &\to& {1 \over 2}{u (\mu - 1) \over 1-u} \cr 
\alpha_2 &\to& -{\sqrt{1 - u} \over \sqrt{e}}\,.
\eeas
Then, defining
\be
{\cal I}_0 \equiv {4^{1 \over d} \over 4 \pi} \int_1^\infty {dy \over y^{{1 \over 2} + {1 \over d}} \sqrt{y-1}} = {4^{1 \over d} \over 4 \pi} B \left({1 \over 2},{1 \over d}\right) = {4^{1 \over d} \over 4 \pi}{\Gamma\left({1 \over 2}\right) \Gamma \left({1 \over d} \right) \over \Gamma \left( {1 \over 2} + {1 \over d} \right)}
\ee
we get from (\ref{resR1b})
\be
R_1 = {2 \Delta z_1 \over \beta_1} \approx {{\cal I}_0 \over e^{{1 \over 2} - {1 \over d}}} {\sqrt{1-u} \over (1 - \mu u)^{1 \over d} }
\ee
and using (\ref{resR1a})
\be
R_2 = R_1\left({1 \over u}, {1 \over \mu},e\right) \approx 1 -\mu^{1 \over d} u^{{2 \over d} -1} {{\cal I}_0  \over e^{{1 \over 2} - {1 \over d}}} {\sqrt{1-u} \over (1 - \mu u)^{1 \over d} }\,.
\ee
Using (\ref{resx}), we find that the relationship between $x$ and $\mu$ to leading order in $e$ is
\be
\mu(x) = {e^{{d \over 2} - 1} \over u^2} \left({ u \over (1 +x) \sqrt{1-u} {\cal I}_0} \right)^d
\ee
From this, we find that the expressions for $E_1 = R_1$  and $E_2 = R_2$ at leading order in $e$ are 
\beas
E_1 &\approx& {1 \over e^{{1 \over 2} - {1 \over d}}} {\cal I}_0 \sqrt{1-u} \cr
E_2 &\approx& {x \over 1 + x}
\eeas
We note that $E_1$ is independent of $x$ at leading order in $e$; the leading $x$ dependence for $E_1$ may be obtained by multiplying the leading order expression by 
$(1 + {1 \over d} u \mu(x))$.

Thus, we see that for $d>2$, $E_1 = R_1$ can be made arbitrarily large by taking $e$ small. In this case, $E_1$ varies very little as a function of $x$, while $R_2$ varies from 0 to 1 as $x$ varies from 0 to $\infty$.

\subsection{Extension to defect CFTs and layered CFTs}

We note that the enhancement of negative vacuum energies that we have observed persists in the limits $x \to 0$ and $x \to \infty$. 

In the $x \to 0$ case, the width of the CFT${}_2$ strip goes to zero, so the layer of CFT${}_2$ becomes a defect in CFT${}_1$. As pointed out in \cite{Bachas:2021fqo}, there are cases when this defect is non-trivial, such that in the holographic dual we still have a wedge of the gravity dual of CFT${}_2$ emerging from the asymptotic boundary (similar to \cite{Akal:2020wfl}) which closes off in the interior to form a bubble. This corresponds to the two boundaries of the bulk interface on the right in Figure \ref{fig:glued}b coming together. 

In the $x \to \infty$ case, we end up with a layer of CFT${}_1$ inside an infinite bulk of CFT${}_2$ on $\mathbb{R}^{d-1,1}$. In this limit, $\mu_2 \to 0$ and the bulk region dual to CFT${}_2$ is locally pure AdS.  
Thus, periodicity is not required for the negative energy enhancement that we observe. 

\section{Field theory toy model}

Let us try to understand the field theory mechanism that allows enhanced negative vacuum energies. We have found that for CFT${}_1$ on a strip of width $w$, coupling the two sides via another CFT with lower central charge can lead to negative vacuum energies much larger in magnitude than for CFT${}_1$ on a periodic direction of width $w$. 

A significant feature of the dual geometries in this case is that the region associated with CFT${}_1$ is a thin strip that wraps the AdS soliton geometry multiple times, with $r_0 \gg r_H$. Since the middle part of the geometry is removed, it would appear that the infrared modes of the field theory are suppressed somehow by the interaction with CFT${}_2$. 
We note that the (artificial) suppression of infrared modes also leads to large negative energies in simple free-field theory toy models.

Consider first a free scalar field in 1+1 dimensions on a periodic direction of length $L$. The Casimir energy density may be calculated by introducing a UV regulator,  calculating the regularized energy density with the periodic direction of length $L$, subtracting the regularized energy density for $L \to \infty$, and then removing the UV regulator (for details, see for example \cite{Bordag:2001qi}). The result is
\be
T_{00} = - {\pi \over 6 L^2} \; . 
\ee
This includes contributions from the individual modes with wave number $2 \pi n /L$. Now, suppose we artificially remove the modes with $|n| \le k$. These provided a finite contribution to the vacuum energy density, 
\beas
T^{|n| \le k}_{00} &=& {1 \over L} \sum_{n=-k}^k {1 \over 2} \omega_n \cr
&=& {1 \over L}\sum_{n=-k}^k {1 \over 2} {2 \pi |n| \over L} \cr
&=& {\pi k (k+1) \over L^2} \,.
\eeas
Thus, the total Casimir energy with these modes removed would be
\be
T_{00} = - {\pi \over L^2} (k(k+1) + {1 \over 6}) \; . 
\ee
In terms of the scale $\Lambda_{IR} = k/L$ associated with the minimum wavelength, this gives
\be
T_{00} = - \pi \left(\Lambda_{IR}^2 + {\Lambda_{IR} \over L} + {1 \over 6 L^2}\right) \; . 
\ee

Similar results are obtained for the Casimir energy of a higher-dimensional free scalar on $S^1 \times R^{d-2,1}$ if we remove modes with small $\omega$. Here, it is convenient to temporarily take the infinite spatial directions to be compact, each with length $A$. Letting $n$ and $\vec{n}$ be the mode numbers in the original periodic direction and the other spatial directions, we consider eliminating modes whose wave numbers satisfy
\be
\sqrt{ \left({n \over L} \right)^2  + \left({\vec{n} \over A}\right)^2 } < \Lambda_{IR}\,.
\ee
The contribution of these modes to the Casimir energy is
\be
T^{IR}_{00} = {1 \over L A^{d-2}} {1 \over 2} \sum_{IR} \sqrt{\left({2 \pi n \over L}\right)^2 +\left({2 \pi \vec{n} \over A}\right)^2 }\,.
\ee
Assuming that $\Lambda_{IR} \gg 1/L$, we can approximate the sum by an integral to obtain
\be
T^{IR}_{00} \approx {1 \over 2  (2 \pi)^{d-1}} \int_{|\vec{k}| < 2 \pi \Lambda_{IR}}  d^{d-1} k  |\vec{k}| = {\pi \Omega_{d-2} \over 2  d} \Lambda_{IR}^d\,,
\ee
so the Casimir energy with the IR modes removed behaves as
\be
T_{00} \approx - {\pi \Omega_{d-2} \over 2  d} \Lambda_{IR}^d
\ee
with corrections controlled by powers of $1 /(L \Lambda_{IR})$.

Recalling that the usual Casimir energy density in this case is of order\footnote{Specifically, we have $- \zeta(3)/(2 \pi L^3)$ for $d=3$ and $-\pi^2/(90 L^4)$ for $d = 4$.} $-1/L^d$ we see that removing the IR modes leads to a large enhancement by a factor of order $(L \Lambda_{IR})^d$.

In these free field toy models, we have simply removed the IR modes by hand. In these cases, it seems unlikely that there is some way to couple the two sides of a CFT on a strip that has the effect of removing these modes. In the holographic models, this suppression of the IR modes happens dynamically, since the dual geometries show that the interior region of the geometry dual to CFT${}_1$ is absent. 

In the end, it's not clear whether the field theory effect pointed out here is relevant to the enhanced negative energies in the strongly coupled holographic models. In particular, the free field theory calculations behave similarly in $d=2$ and $d>2$, while in the holographic model, the large negative energies appear only for $d>2$. 

\section{Discussion}
  
In this note, we have made use of a holographic model for interface CFTs to demonstrate a surprising enhancement of negative vacuum energy for a CFT on a strip. This effect occurs when both sides of the strip are coupled to a CFT with fewer degrees of freedom, and the interfaces are of the right type, corresponding to a bulk interface tension close to a critical value.

In the context of the holographic model, we can make the magnitude of the negative energy density arbitrarily large for a strip of fixed width by tuning the bulk interface tension arbitrarily close to a lower critical value. For specific microscopic CFTs\footnote{See \cite{Chiodaroli:2011fn, Chiodaroli:2012vc, DHoker:2007zhm, DHoker:2007hhe, Aharony:2011yc,Assel:2011xz} for top-down examples of holographic interface CFTs.}, we may not have a continuously variable parameter associated with the interface, so it is interesting to understand whether there exist interfaces that have these qualitative properties. We note that the bulk interface tension is positively related to a central charge associated with the interface, which is expected to decrease under renormalization group flows (see e.g. \cite{Estes:2014hka}). Thus, the interfaces that give rise to the more negative energies may be expected to be the ones with minimal values of this central charge that have no relevant perturbations. An interesting class of microscopic models to study would be those where the bulk CFTs are ${\cal N}=4$ supersymmetric Yang-Mills theory with different gauge group rank and/or gauge coupling, and the defects correspond three-dimensional SCFTs \cite{Assel:2011xz} which individually preserve half the original supersymmetry but together break the supersymmetry. As discussed in \cite{VanRaamsdonk:2021qgv}, despite the absence of supersymmetry, these models may still have a weakly curved dual description in terms of type IIB supergravity.
  
It would be very interesting if the negative energy enhancement we observe here is relevant to some real-world systems with a layer of one material inside another material, either for two or three dimensional materials. While our modes are holographic, with a large number of local degrees of freedom, it is not clear that this large number of local degrees of freedom is essential for the effect. 

The appearance of large negative energies may be particularly interesting in a gravitational context where the negative energy can be used to support various exotic geometries. For example, as discussed in \cite{VanRaamsdonk:2021qgv}, taking the setup of Figure \ref{fig:setup} and turning on gravity in the CFT$_1$ region the semiclassical Einstein equations have eternally traversable wormhole solutions when the interface is tuned appropriately.

\subsubsection*{Anomalously high transition temperatures in layered materials?}

We have focused on the vacuum properties, but it would be interesting to understand more completely the physics at finite temperature, generalizing the results of \cite{Bachas:2021fqo} to higher dimensions. In particular, it would be interesting to see how stable the solutions with enhanced negative energies are to finite temperature perturbations.

For a holographic CFT with one periodic direction (with antiperiodic boundary conditions for fermions), we have a phase transition when the inverse temperature equals the length of the periodic direction \cite{Witten1998a}. Above this temperature, the thermal circle is contractible in the bulk. The CFT with a compact direction can be understood as a confining gauge theory from the lower-dimensional point of view, and the phase transition is a deconfinement transition. In the layered situation, we similarly expect a phase transition to a high temperature phase where the thermal circle becomes contractible. The naive scale for this transition would be the width of the CFT$_1$ layer. However, we found that the bulk solution corresponding to a CFT$_1$ layer can be that corresponding to the CFT with a periodic direction whose length is much smaller than the width of the interval. In this case, we expect that this higher temperature scale sets the transition temperature.

\begin{figure}
    \centering
    \includegraphics[width=0.25\textwidth]{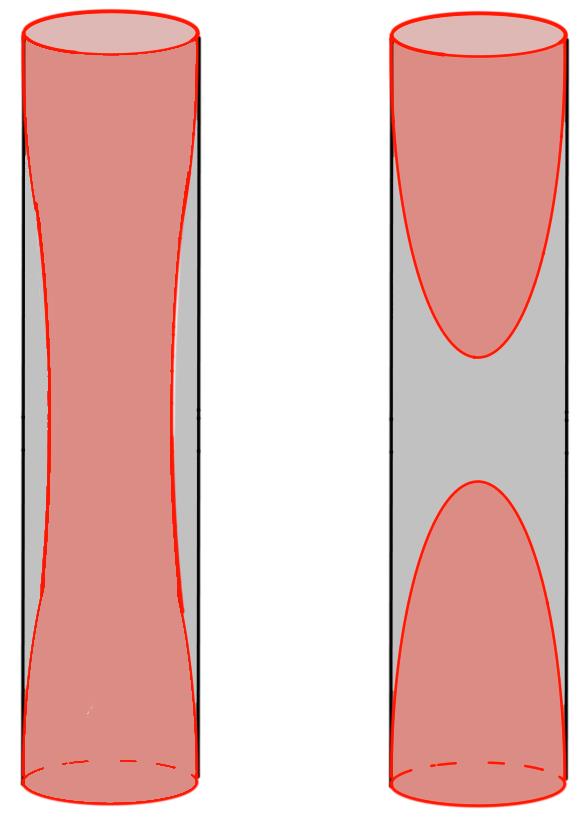}
    \caption{Bulk region associated with CFT${}_1$ (shown in grey) at low and high temperature. As the interface tension approaches the lower critical value, we expect that the transition temperature must be increasingly large in order to admit the high-temperature phase disconnected solution (right)  which avoids intersection of the domain walls.}
    \label{fig:highT}
\end{figure}

To see this, we note that during the transition in the interface case, the bulk interfaces associated with the two CFT interfaces become disconnected and the thermal circle is contractible in the bulk, as shown in Figure \ref{fig:highT}. In the limit where the interface tension approaches the lower critical value (where the local state of CFT$_1$ is similar to that of a CFT on a circle much smaller than the width of the strip), the angle of intersection between the interface and the AdS boundary becomes small. In this case, we expect that the upper interface in the right side of Figure \ref{fig:highT} extends further and further down from the interface. Having a solution as in the right  of Figure \ref{fig:highT} which avoids intersection of the two interfaces would require an increasingly large temperature (small periodic direction) relative to the width of the interval (vertical direction) as the tension is lowered to the critical value. Thus, we expect that the transition temperature will become large in this limit. If this is correct, it means that the large negative energy density in the CFT${}_1$ region will persist to relatively high temperatures. 

It would be interesting to explore whether a similar mechanism could lead to anomalously high transition temperatures for holographic layered materials exhibiting other low-temperature phases, e.g. those exhibiting superconductivity. 
There is a wealth of fascinating phenomena associated with layered materials in real-world condensed matter systems (e.g. high-T$_c$ cuprate superconductors). It would be nice to understand whether the underlying field theory mechanism responsible for the anomalously large vacuum energies and anomalously high transition temperatures (assuming our qualitative arguments for the latter are correct) in our holographic model is relevant to observable phenomena in experimentally realizable layered materials. 

\section*{Acknowledgements}

We would like to thank Costas Bachas, Ben Freivogel, Henry Lin, and Juan Maldacena for discussions. MVR is supported by the Simons Foundation via the It From Qubit Collaboration and a Simons Investigator Award. PS is supported by the Natural Sciences and Engineering Research Council of Canada via the Canada Graduate Scholarship. AM is supported by a C-GSM award given by the National Science and Engineering Research Council. 

\appendix

\section{Other possible vacua for the CFT on $\mathbb{R}^{d-2,1} \times S^1$?}

\begin{figure}
    \centering
    \includegraphics[width=100mm]{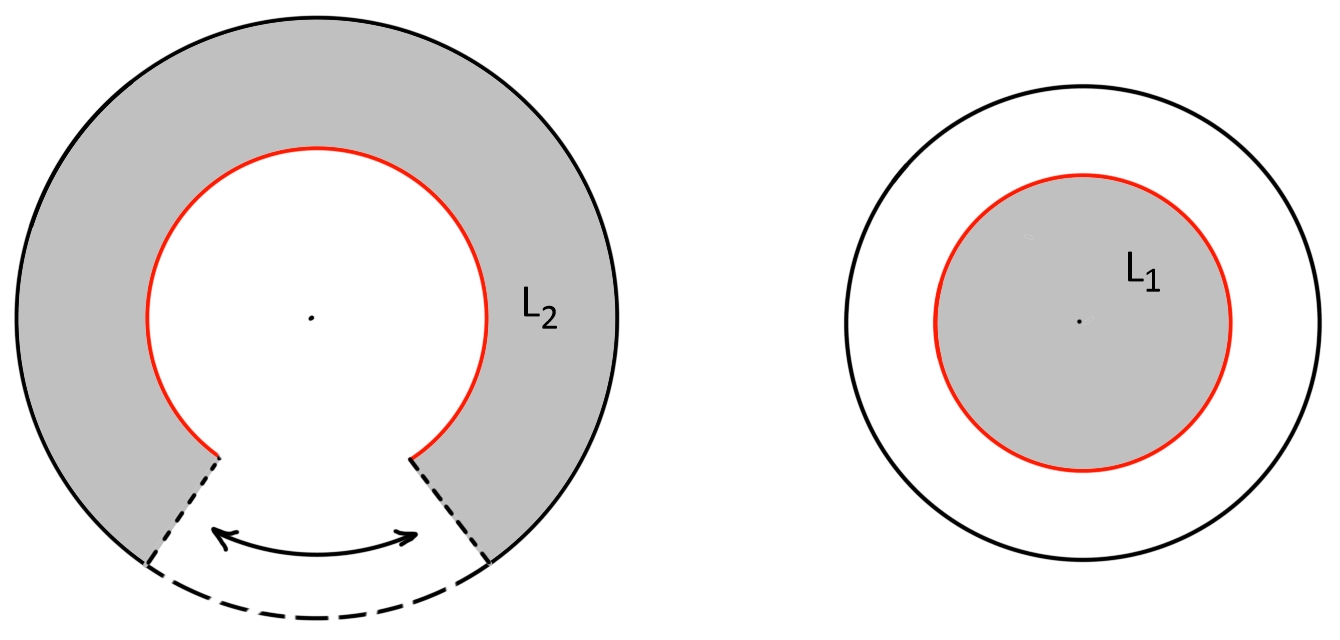}
    \caption{Possible geometry dual to a CFT on $R^{d-2,1} \times S^1$ with an interior bubble described by a different low-energy effective theory (see appendix A).}
    \label{fig:bubble}
\end{figure}

Since the non-perturbative gravitational theory dual to a CFT must include domain walls associated with allowed interfaces to other CFTs \cite{Simidzija:2020ukv}, an interesting possibility is that we could have Poincar\'e invariant states of a holographic CFT on $R^{d-2,1} \times S^1$ dual to a geometry with a domain wall, as shown in Figure \ref{fig:bubble}. This would include a region $r<r_0$ of the geometry (\ref{geom}) with AdS length $L_1$ glued into a region $r > r_0$ of the geometry (\ref{geom}) with AdS length $L_2$. For the interior geometry, the periodicity of the $z$ direction must be (\ref{defbeta}) in order to avoid a conical singularity, but for the exterior geometry, we can have any period.

In this case, the extrinsic curvatures measured with respect to region $i$ using a normal vector pointing towards the outer region are
\be
K_{zz} = {L_i^2 \over 2}{\partial f_i \over \partial r} \sqrt{f_i(r)} \qquad K_{\mu \nu} = \eta_{\mu \nu} r \sqrt{f_i(r)} \; ,
\ee
so the junction conditions (\ref{eq:JC2}) give
\bea
r \sqrt{f_1(r)} - r \sqrt{f_2(r)} &=& \kappa r^2 \cr
{1 \over \sqrt{f_1}} {\partial f_1 \over \partial r} - {1 \over \sqrt{f_2}} {\partial f_2 \over \partial r} &=& 2 \kappa \; .
\eea

Thus, we can have a solution for some radius $r_0$ if $v(r_0) = v'(r_0) = 0$, where
\be
v(r) \equiv \sqrt{f_1(r)} - \sqrt{f_2(r)} - \kappa r \; .
\ee
From $v(r_0)=0$, we conclude that the tension must be related to $r_0$ and the parameters $L_1$ and $L_2$ by
\be
\label{eqkappa}
\kappa = {1 \over r_0} \left(\sqrt{f_1} - \sqrt{f_2}\right) \; .
\ee
The remaining equation gives
\be
\left. {d \over dr} \left({\sqrt{f_1(r)} \over r} - {\sqrt{f_2(r)} \over r} \right) \right|_{r = r_0} = 0
\ee
Solving this, we find 
\be
r_0^d = \mu_1 \mu_2 {\mu_1 - \mu_2 \over \left({\mu_1 \over L_2}\right)^2 - \left({\mu_2 \over L_1}\right)^2} \; .
\ee
We must have that $r_0$ is larger than the minimum value $r_H$ (Equation \ref{defrh}) for each region. Defining as before $\mu = \mu_2/\mu_1$ and $u = L_2/L_1$, we must have one of the following options:
\be
u > 1 \qquad {1 \over u^2} < \mu < {1 \over u} \qquad \qquad  OR \qquad \qquad u < 1
 \qquad {1 \over u} < \mu < {1 \over u^2}
\ee
From (\ref{eqkappa}), we have that
\be
\kappa L_1 = {1 \over u} \sqrt{1 - \mu u^2 \over \mu - 1} \left({1 \over \sqrt{\mu}} - \sqrt{\mu}\right)
\ee
so we have a positive tension domain wall only for the $u > 1$ case. Thus, the allowed region of parameter space for bubble solutions with positive tension domain walls is
\be
\label{murange}
u > 1 \qquad \qquad {1 \over u^2} < \mu < {1 \over u} \; .
\ee
We find that $\kappa$ increases from $0$ to $\kappa_-$ as $\mu$ increases from $1/u^2$ to $1/u$. Thus, the tension is always less than the lower critical value $\kappa_-$. Therefore, these domain walls cannot reach the AdS boundary; they are not of the type that corresponds to an interface between two CFTs. 

Assuming that such domain walls exist in some cases, we can still consider the energy associated with these solutions. To avoid a conical singularity, the periodicity of $z_1$ must be given by $\beta_1$ in (\ref{defbeta}). The proper distance around the domain wall must match in the inner and outer spacetimes, so the periodicity of $z_2$ must satisfy
\be
L_2 \sqrt{f_2} \Delta z_2 = L_1 \sqrt{f_1} \beta_1 \; .
\ee
Finally, we have that the ratio of the periodicity of $z_2$ to the periodicity $\beta_2$ in the solution without an interface is
\be
\label{Rsimp}
R = {\Delta z_2 \over \beta_2} = {L_1 \sqrt{f_1} \beta_1 \over L_2 \sqrt{f_2} \beta_2}  = (\mu u^2)^{1/d-1}\; .
\ee
We recall that $R^d$ gives the ratio of the energy density to that for the case where the bulk solution is just the usual AdS-soliton. So we would have a lower-energy vacuum if there are any cases with $R>1$. But the constraints (\ref{murange}) imply that $\mu u^2 >1$, so for positive tension branes, we only get $R<1$. 

We conclude that considering solutions with a bubble does not lead to a lower vacuum energy than we had from the AdS soliton solution.

We have also checked that these solutions are not perturbatively stable. Starting with one of the solutions, we consider keeping the outer geometry fixed while varying the $r$ coordinate of the interface. We continue to impose the first junction condition so that the inner and outer regions join without a discontinuity in the metric. This requires that the parameter $\mu_1$ be adjusted along with the interface radius in a particular way. Evaluating the action for the resulting off-shell configuration, we find that in all cases, our solution represents a local maximum as a function of the interface radius.
  
\bibliographystyle{jhep}
\bibliography{references}

\end{document}